\documentclass[aps,pre,twocolumn,superscriptaddress,showpacs]{revtex4-1}
\usepackage{amssymb, amsmath}
\usepackage{color}
\usepackage{graphicx}

\begin{document}

\title{Diffusion anomalies in ac driven Brownian ratchets}
\author{Jakub Spiechowicz}
\affiliation{Institute of Physics, University of Silesia, 40-007 Katowice, Poland}

\author{Jerzy {\L}uczka}
\affiliation{Institute of Physics, University of Silesia, 40-007 Katowice, Poland}
\affiliation{Silesian Center for Education and Interdisciplinary Research, University of Silesia, 41-500 Chorz{\'o}w, Poland}
\email{jerzy.luczka@us.edu.pl}

%
\begin{abstract}
We study diffusion in ratchet systems. As a particular experimental realization we consider an asymmetric SQUID subjected to an external ac current and a constant magnetic flux. We analyze mean-square displacement of the Josephson phase and find that within selected parameter regimes it evolves in three distinct stages: initially as superdiffusion, next as subdiffusion and finally as normal diffusion in the asymptotic long-time limit. We show how crossover times that separates these stages can be \emph{controlled by temperature and an external magnetic flux}. The first two stages can last many orders longer than characteristic time scales of the system thus being comfortably detectable experimentally. The origin of abnormal behavior is noticeable related to the ratchet form of the potential revealing an \emph{entirely new mechanism of emergence of anomalous diffusion}. Moreover, a normal diffusion coefficient exhibits non-monotonic dependence on temperature leading to an intriguing phenomenon of \emph{thermal noise suppressed diffusion}. The proposed setup for experimental verification of our findings provides a  new and  promising \emph{testing ground} for investigating anomalies in diffusion phenomena.
\end{abstract}
\maketitle

\section{Introduction}

The theory of Brownian motion has played a guiding role in the development of  statistical physics. It provides a link between the microscopic dynamics and the observable macroscopic phenomena such as diffusion. The latter has been in the research spotlight already for over 100 years \cite{hanggi100years}. One century after pioneering Einstein's work it remains both a fundamental open issue and a continuous source of developments for many areas of science. Recent prominent examples include stochastic resonance \cite{gammaitoni1998}, ratchet effects \cite{hanggi2009}, enhancement of diffusion \cite{reimann2001} or efficiency \cite{spiechowicz2014pre}, to name but a few. Due to universal character of diffusion and its ubiquitous presence both in classical 
and quantum 
systems as well as in molecular biology it is still a subject of intensive studies. In last years its anomalous character has been one of the main research topic in various fields \cite{metzler2000, klafter2005, metzler2014, zaburdaev2015}. In particular, diffusion of Brownian particles in deterministic periodic or  random potentials,  or a combination of both has been investigated \cite{sancho2004,khoury2011,sune2013,simon2014}. Anomalous behavior may not survive until the asymptotic long time regime nonetheless lately its transient nature has been predicted theoretically and observed experimentally \cite{bronstein2009,hanes2012theo,hanes2012exp,hanes2013}. In this work we study a model which belongs to an archetypal class of Brownian ratchets \cite{hanggi2009}. Despite its simplicity it is able to exhibit an extremely rich dynamics and variety of \emph{anomalous transport} features like the absolute negative mobility in a linear response regime \cite{machura2007,speer2007,nagel2008}, the negative mobility in a nonlinear response regime and the negative differential mobility \cite{kostur2008}. However, here we focus on diffusion anomalies occurring in this system.
\begin{figure}[b]
	\centering
	\includegraphics[width=0.88\linewidth]{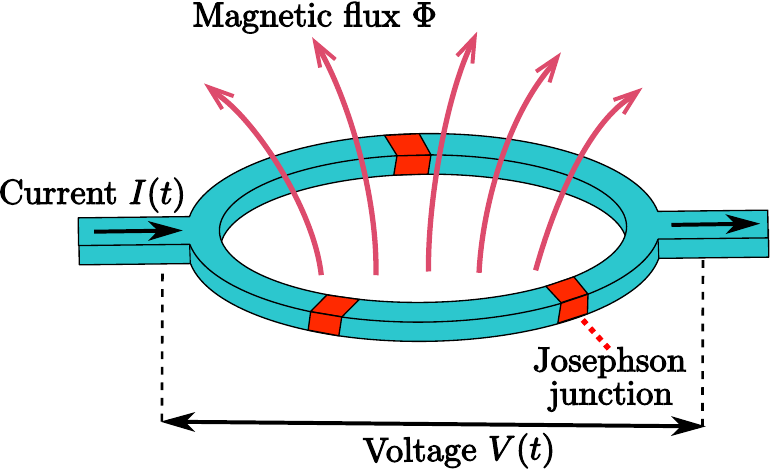}
	\caption{The asymmetric SQUID composed of three Josephson junctions and driven by the external current $I(t)$. The external constant magnetic flux is $\Phi$ and the instantaneous voltage across the SQUID is $V(t)$.}
	\label{fig1}
\end{figure}

Our model consists of four relevant components and can be formulated in terms of (i) a classical inertial Brownian particle of mass $M$, (ii) moving in a deterministic ratchet potential $U(x)=U(x+L)$ of period $L$, (iii) driven by an unbiased time-periodic force $a\cos{(\omega t)}$ of amplitude $a$ and angular frequency $\omega$, and (iv) affected by thermal noise of temperature $T$. There are  examples  of experimentally accessible physical systems that can be described by this type of a model. An important representative that comes to mind is transport of ions through nanopores \cite{karnik2007}, cold atoms in optical lattices \cite{gommers2005, arzola2011, denisov2014}, type II superconducting devices based on motion of Abrikosov vortices \cite{lee1999, villegas2003}, Josephson vortices \cite{ustinov2004, beck2005, knufinke2012} and a superconducting phase in weak links and SQUIDs \cite{weiss2000, sterck2005, sterck2009}, to give only a few. To maintain a close link with recent experimental research in this field and challenge experimentalists to put our theoretical predictions into a reality check from now on we stick to a particular realization of a rocking ratchet mechanism, namely the asymmetric SQUID \cite{zapata1996,spiechowicz2014}. This Josephson-Brownian ratchet offers some advantages over other setups: (i) precise experimental control of applied driving forces here in the form of external currents, (ii)  detection of directed motion manifested in a non-zero long-time dc voltage, (iii) access to studies over a wide frequency range of adiabatic and non-adiabatic external perturbations and finally (iv) both underdamped and overdamped dynamics can be investigated by proper junction fabrication and variation of system parameters. We want to emphasize that our findings are \emph{universal} in the sense that they apply to a broad selection of physical setups and could be observed in variety of experimental realizations of a rocking ratchet mechanism. In this context, 
we recommend readers refer to the papers  \cite{renz12,renz13}, where normal and anomalous diffusion  in ac driven systems of cold atoms in dissipative optical lattices  has been studied both experimentally and theoretically. Moreover, in Ref. \cite{renz13}, the survey of previous experiments on anomalous diffusion in such systems is presented. 

Our work is organized as follows. In Sec. II we introduce the model and all quantities of interest. Sec. III contains a detailed description of our results, in particu- lar control of anomalous diffusion by temperature and an external magnetic field as well as explanation of the mechanism that stands behind observed diffusion anomalies. In Sec. IV we discuss temperature dependence of the diffusion coefficient. Finally, Sec. V is devoted to summary and conclusions. 
\section{Model}
\label{model}
As an exemplary real system we consider a SQUID presented in Fig. \ref{fig1}. It is a loop with three resistively and capacitively shunted Josephson junctions: two identical   are placed in one arm whereas the third  is located in the other arm. Additionally, the SQUID is threaded by an external constant magnetic flux $\Phi$ and driven by a time periodic current $I(t)$. There is a  one-to-one correspondence between this setup and a classical Brownian particle. The particle position $x$ translates to the phase $\varphi=\varphi_1+\varphi_2$, where $\varphi_1$ and $\varphi_2$ are the Josephson phases of two junctions located in the same arm. The particle velocity $v=\dot{x}$ translates to the voltage $V \propto \dot \varphi$ across the SQUID, the external force to the current $I(t)$, the particle mass $M$ to the capacitance $M \propto C$ and the friction coefficient $\gamma$ to the normal conductance $\gamma \propto G=1/R$. Here we present only the dimensionless form of the Langevin equation governing the phase dynamics. For details we refer the reader to our recent paper \cite{spiechowicz2014}. It reads
\begin{equation}
    \label{eq1}
    \tilde{C} \ddot{x}(t) + \dot{x}(t) = -U'(x(t)) + a\cos(\omega t) + \sqrt{2Q}\,\xi(t), 
\end{equation}
where the dot and prime denotes a differentiation with respect to the dimensionless time $t$ and the rescaled phase $x =(\varphi+\pi)/2$, respectively. Other dimensionless quantities appearing in this formula are: the  capacitance  $\tilde{C}$ of the device, the amplitude $a$ and the frequency $\omega$ of the external ac current $I(t)$. Johnson-Nyquist thermal noise is modelled by  symmetric and unbiased $\delta$-correlated Gaussian white noise $\xi(t)$ of average $\langle \xi(t) \rangle = 0$ and the correlation function $\langle \xi(t)\xi(s) \rangle = \delta(t-s)$. Its intensity \mbox{$Q \propto k_B T$} is proportional to thermal energy, where $T$ and $k_B$ is the system temperature and the Boltzmann constant, respectively. The spatially periodic potential $U(x)$ of period $2\pi$ is in the following form  \cite{spiechowicz2014}
\begin{figure}[t]
    \centering
    \includegraphics[width=0.88\linewidth]{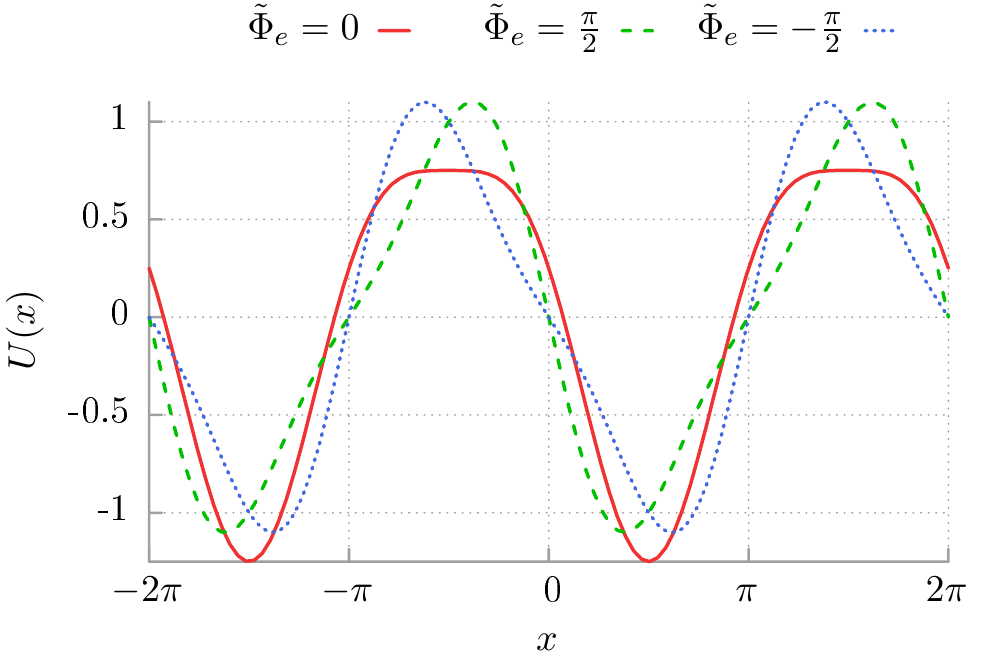}
    \caption{The potential (\ref{eq2}) for $j=1/2$ in the symmetric case $\tilde\Phi_e =0$  (solid red line) in comparison with the ratchet potential for two values of the external magnetic flux $\tilde\Phi_e = \pi/2$ (dashed green line) and $\tilde\Phi_e = -\pi/2$ (dotted blue line).}
    \label{fig2}
\end{figure}
\begin{equation}
    \label{eq2}
    U(x) = - \sin(x) - \frac{j}{2} \sin(2x + \tilde\Phi_e - \pi/2).
\end{equation}
The parameter $j=J_2/J_1$ is a ratio of critical currents of two junctions in opposite arms and $\tilde{\Phi}_e$ is the dimensionless external constant magnetic flux. If $j \neq 0$ the potential is generally asymmetric and its reflection symmetry is broken, see Fig. \ref{fig2}. However, even when $j \neq 0$ there are certain values of the external magnetic flux $\tilde{\Phi}_e$ for which it is still symmetric.
\begin{figure*}[t]
    \centering
    \includegraphics[width=0.4\linewidth]{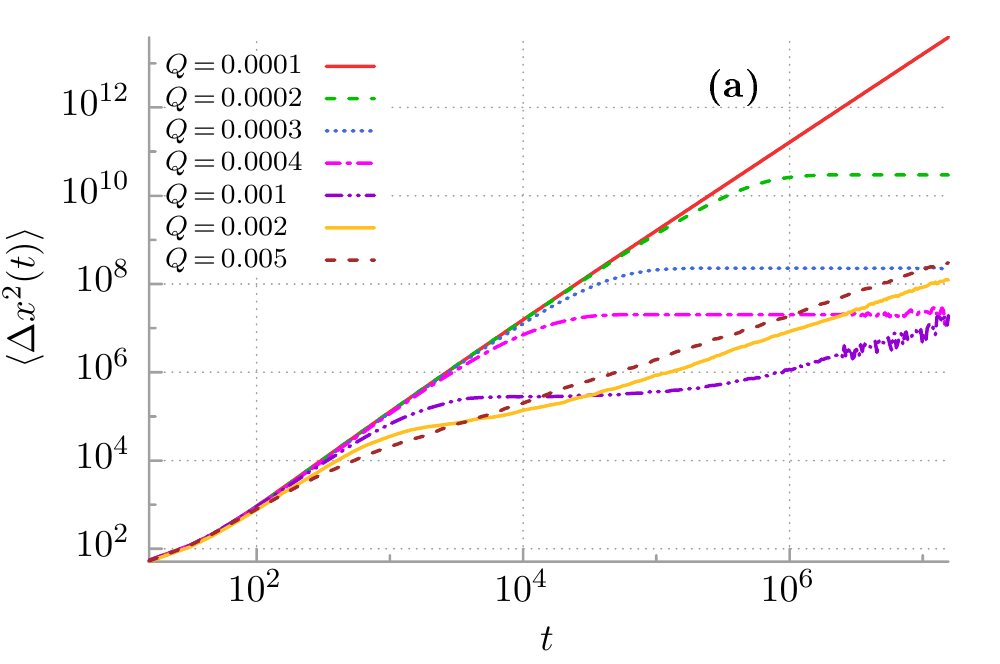}
    \includegraphics[width=0.4\linewidth]{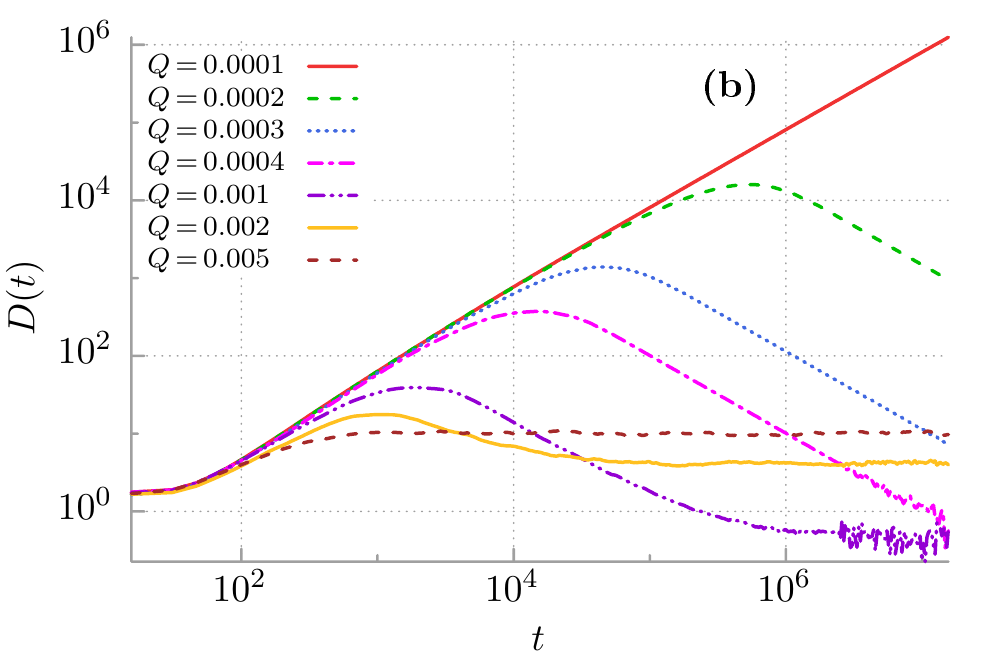} \\
    \includegraphics[width=0.4\linewidth]{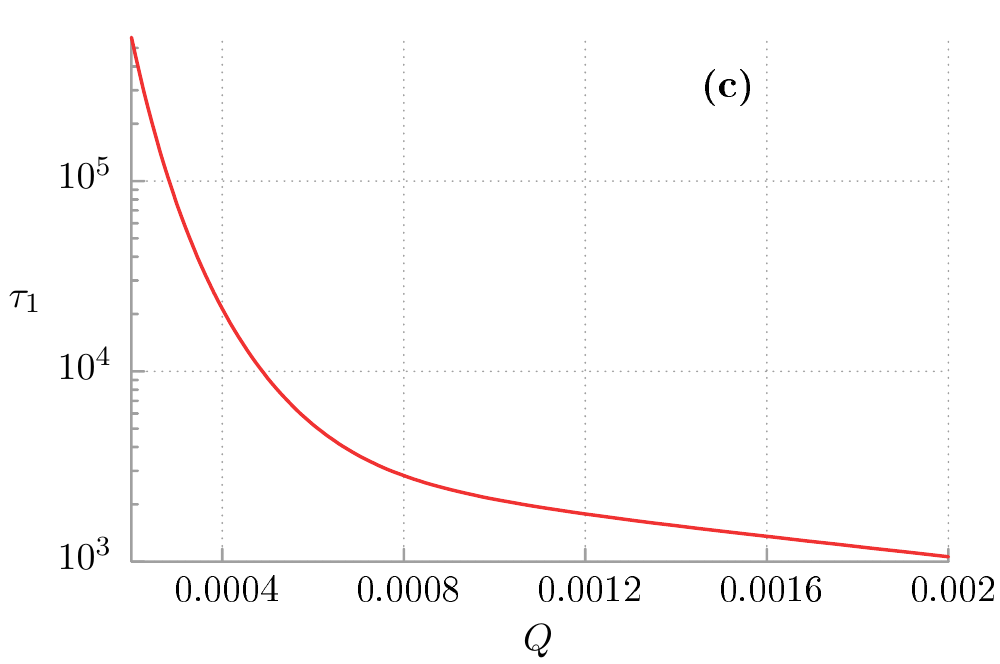}
    \caption{Impact of temperature $Q \propto T$ on the diffusion process.  (a) The mean square displacement  $\langle \Delta x^2(t)\rangle$ of the Josephson phase. (b) The diffusion coefficient $D(t)$. (c) The crossover time $\tau_1$ separating superdiffusion and subdiffusion  stages. Parameters are: $\tilde{C} = 6$, $a = 1.899$, $\omega = 0.403$,$\tilde{\Phi}_e = \pi/2$ and $j = 0.5$.}
    \label{fig3}
\end{figure*}

The most important quantity characterizing diffusion of the phase $x(t)$ is its mean square displacement (MSD) $\langle \Delta x^2(t) \rangle$ defined as
\begin{equation}
    \label{eq3}
    \langle \Delta x^2(t) \rangle = \left\langle [ x(t) - \langle x(t) \rangle ]^2 \right\rangle,
\end{equation}
where $\langle \cdot \rangle$ indicates an average over initial conditions and thermal noise realizations. Although the dynamics may not be normal diffusion at all times nonetheless a time-dependent diffusion coefficient $D(t)$ can be defined as
\begin{equation}
    \label{eq4}
    D(t) =  \frac{\langle \Delta x^2(t) \rangle}{2t}.
\end{equation}
Information on the diffusive process is also contained in the slope of the MSD which can be obtained from the power-law fitting \cite{atbook}
\begin{equation}
    \label{eq5}
    \langle \Delta x^2(t) \rangle \,\sim  t^{\alpha}.
\end{equation}
The exponent $\alpha$ characterizes a type of diffusion. Normal diffusion is for $\alpha =1$. There are two distinct regimes of \emph{anomalous diffusion}: subdiffusion for $0 < \alpha < 1$ and superdiffusion for $\alpha > 1$. 
In the former case the MSD increases over time slower and in the latter case faster than the rate of normal diffusion. Another special case is ballistic diffusion when $\alpha = 2$. When the value of $\alpha$ is guaranteed to be unity, the time-independent diffusion coefficient $D$ can be  determined as
\begin{equation}
    \label{eq6}
    D = \lim_{t \to \infty} D(t). 
\end{equation}
Otherwise the above definition is not constructive because  $D$ is either zero (subdiffusion) or diverges to infinity (superdiffusion). In the case of SQUID, the phase diffusion can be investigated experimentally via measurement of the power spectrum of voltage fluctuations \cite{geisel, ishizaki2003, atbook}.
\section{Results} 
\label{results}
The Fokker-Planck equation corresponding to the Langevin equation (\ref{eq1}) cannot be solved by use of any known analytical methods \cite{jung1993}. Moreover, even in the deterministic limit of vanishing thermal noise intensity $Q = 0$ this system exhibits very complex dynamics including chaotic regimes \cite{jung1996, mateos2000}. Therefore, In order to investigate the  diffusion process  we have carried out comprehensive numerical simulations of the driven Langevin dynamics determined by (\ref{eq1}). All numerical calculations were done by use of a CUDA environment implemented on a modern desktop GPU. This proceeding allowed for a speed-up of a factor of the order of $10^3$ as compared to a common present day CPU method. For up-to-date review of this scheme we refer the reader to \cite{spiechowiczcpc}.
 
The system described by (\ref{eq1}) has a 6-dimensional parameter space $\{ \tilde{C}, a, \omega, j, \tilde{\Phi}_e,  Q\}$ which is too large to analyze numerically in a systematic fashion even with the help of our innovative computational methods. We therefore have decided to focus our attention on the effect of temperature $Q \propto T$ and the impact of the external constant magnetic flux $\tilde{\Phi}_e$ in the remarkable regime of the thermal noise enhanced rectification efficiency of the SQUID studied in detail in Ref. \cite{NJP15}. Unless stated otherwise this case corresponds to the following set of parameters $\{ \tilde{C}, a, \omega, j, \tilde{\Phi}_e, Q\} = \{ 6, 1.899, 0.403,  0.5, \pi/2, 0.0004\}$.
\begin{figure*}[t]
    \centering
    \includegraphics[width=0.4\linewidth]{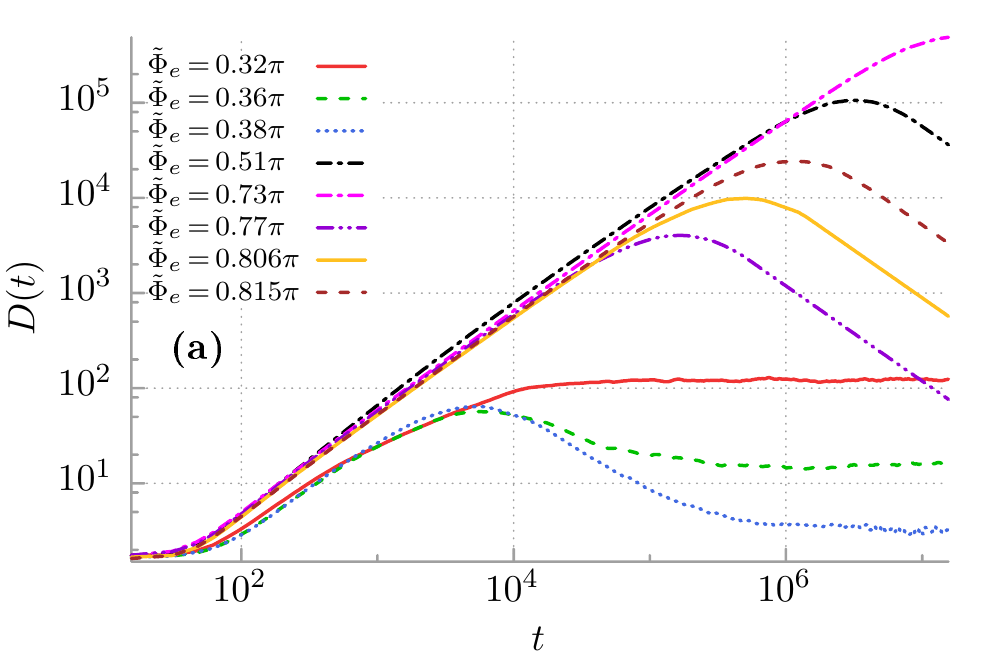}
    \includegraphics[width=0.4\linewidth]{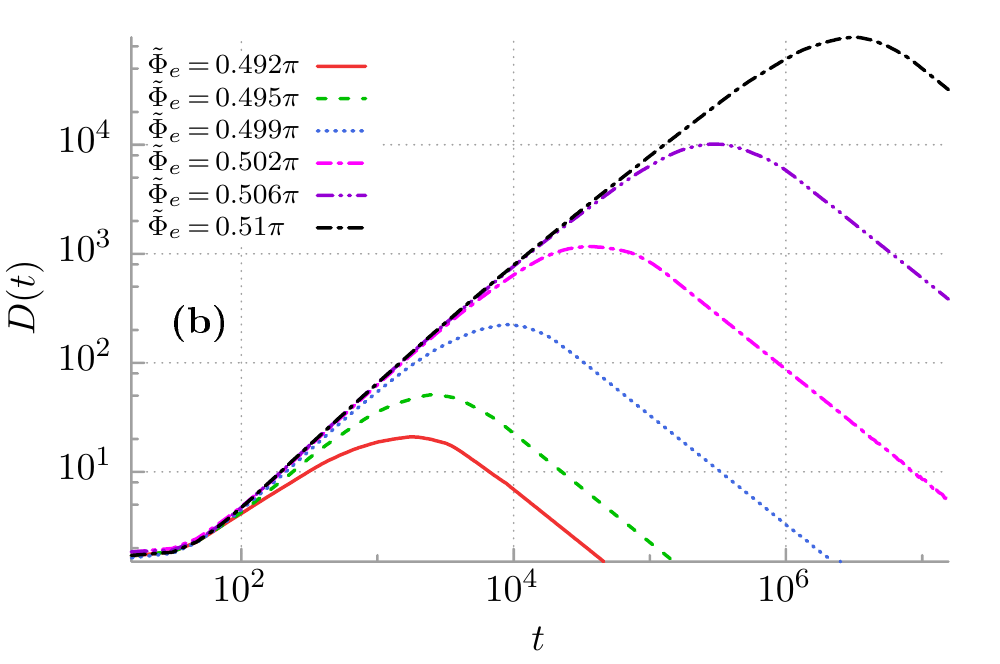} \\
    \includegraphics[width=0.4\linewidth]{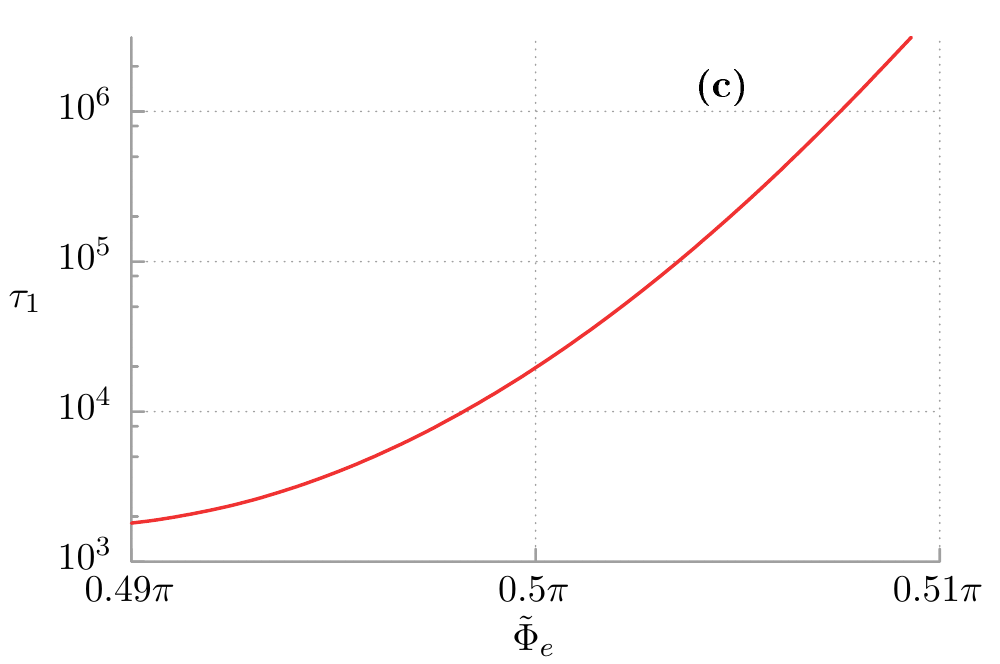}
    \caption{Control of diffusion by the external magnetic flux $\tilde{\Phi}_e$. (a) The diffusion coefficient $D(t)$ for selected values of the flux $\tilde{\Phi}_e$. (b) Sensitivity of the diffusion coefficient $D(t)$  in the vicinity of the magnetic flux $\tilde{\Phi}_e = \pi/2$. (c) The crossover time $\tau_1$ as a function of $\tilde{\Phi}_e$ in the vicinity of $\tilde{\Phi}_e = \pi/2$. Note the colossal increase of order $10^3$ in the crossover time $\tau_1$ when the magnetic flux changes in the small interval $(0.49 \pi, 0.51 \pi)$. Other parameters are the same as in Fig. \ref{fig3} except $Q = 0.0004$.}
    \label{fig4}
\end{figure*}
\subsection{Control of anomalous diffusion regimes  by temperature}
In panel (a) and (b) of Fig. \ref{fig3} we show time evolution of the MSD $\langle \Delta x^2(t) \rangle$ and the diffusion coefficient $D(t)$, respectively, for selected values of the noise intensity $Q \propto T$. Especially in the latter the reader may easily distinguish between the type of diffusion: superdiffusion occurs in the interval where $D(t)$ increases, the case of decreasing $D(t)$ corresponds to subdiffusion and for non-varying $D(t)$ normal diffusion takes place. In the low temperature limit the lifetime of superdiffusion is extremely long, see the case of $Q = 0.0001$ in panel (b). It might lead one incorrectly to conclude that this anomalous diffusion regime occurs in the  stationary regime. This statement is true only in the deterministic case when formally $Q = 0$. However, in such a case the considered SQUID  model is not correct. 
The deflection from the early superdiffusive behavior can expressively be noted as temperature increases. The evolution can be divided into three time-domains: the early period of superdiffusion $\tau_1$, the intermediate interval $\tau_2$ where subdiffusion is developed and the asymptotic long time regime where normal diffusion occurs. The crossover times $\tau_1$ and $\tau_2$ separating these domains can be \emph{controlled by temperature}. For example, when $Q = 0.0004$, $\tau_1 \approx 1.52 \cdot 10^4$ and $\tau_2 \approx 0.9985 \cdot 10^7$. If temperature is increased to $Q = 0.002$, the crossover times are reduced to $\tau_1 \approx 10^3$ and $\tau_2 \approx 0.99 \cdot 10^5$. In panel (c) of the same figure we present the dependence of the crossover time $\tau_1$ separating super- and subdiffusion on temperature $Q \propto T$. It is remarkable that superdiffusion lifetime $\tau_1$ changes nearly three order of magnitude when the thermal noise intensity $Q$ varies in the interval  $(0.0002, 0.002)$. For higher temperatures the phase motion is initially superdiffusive and next normal diffusion takes place, as e.g. in the case $Q = 0.005$ in panel (b) of Fig. 3.
\subsection{Control of anomalous diffusion regimes by external magnetic field}
From the experimental point of view it is more convenient to manipulate transport properties of the SQUID by the external magnetic flux $\tilde{\Phi}_e$. In Fig. \ref{fig4}, its impact on the  diffusion process is illustrated. In some regimes, two crossover times $\tau_1$ and $\tau_2$ are identified and their magnitudes can be \emph{changed by variation of the external magnetic flux} $\tilde{\Phi}_e$. For example, when $\tilde{\Phi}_e = 0.73\pi$, the superdiffusion lifetime is $\tau_1 \approx 1.56 \cdot 10^7$ while for \mbox{$\tilde{\Phi}_e = 0.492\pi$} it is $\tau_1 \approx 1.8 \cdot 10^3$, i.e. four orders shorter. However, in contrast to the case when temperature $Q$ is varied the dependence of the crossover times $\tau_1$ and $\tau_2$ on  $\tilde{\Phi}_e$ is non-monotonic. It can be concluded from  panel (a) of Fig. \ref{fig4}. Moreover, for some intervals of the  magnetic flux these times are extremely sensitive to small changes of $\tilde{\Phi}_e$. This situation is exemplified in panels (b) and (c): small changes of order $10^{-2}$ in $\tilde{\Phi}_e$ are accompanied by the giant increase  of order $10^3$ in the crossover time $\tau_1$. 
For our particular parameter regime the crossover time $\tau_2$ is often so long that its numerical estimation may be controversial due to surely limited stability of the utilized algorithms leading to uncontrolled propagation of round-off and truncation errors \cite{spiechowiczcpc}. Therefore we do not present its dependence on temperature or the external magnetic flux.
\begin{figure}[t]
    \centering
    \includegraphics[width=0.9\linewidth]{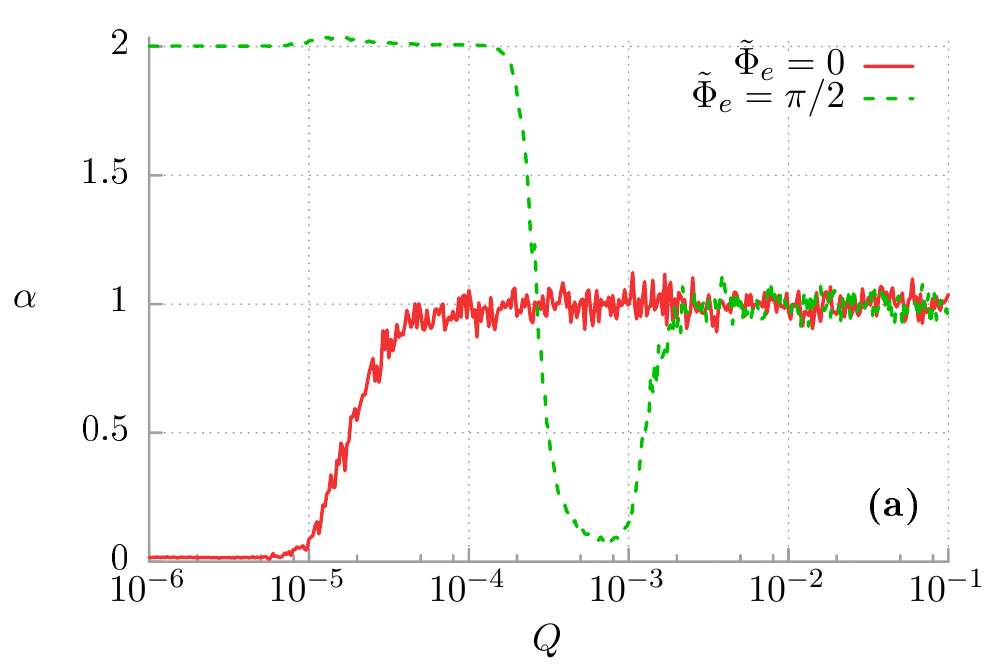}\\
    \includegraphics[width=0.9\linewidth]{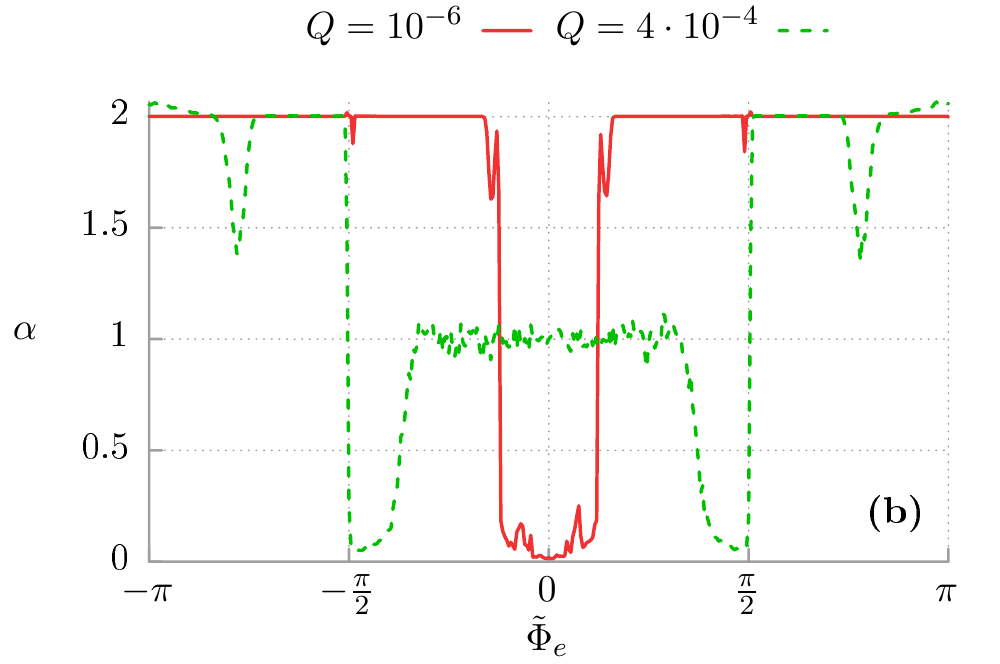}
    \caption{The power exponent $\alpha$ in dependence on temperature $Q \propto T$ and the external magnetic flux $\tilde{\Phi}_e$. Parameters are the same as in Fig. \ref{fig3}.}
    \label{fig5}
\end{figure}
\subsection{Comment on the power exponent}
Let us now ask a complementary question, how the power exponent $\alpha$ defined by the relation (\ref{eq5})  depends on temperature and the external magnetic flux. 
This exponent was fitted from time evolution of the MSD $\langle \Delta x^2(t) \rangle$ over $\approx 10^5$ dimensionless time units at a number of $Q$ and $\tilde{\Phi}_e$ values. The results are presented in Fig. \ref{fig5}. Panel (a) depicts the dependence of $\alpha$ on temperature $Q$ for two different potential profiles $U(x)$, namely \emph{symmetric} ($\tilde{\Phi}_e = 0$) and \emph{ratchet} ($\tilde{\Phi}_e = \pi/2$). In the low temperature limit the crucial role of the potential asymmetry for the emergence of anomalous diffusion is observed. When potential is reflection symmetric then $\alpha \approx 0$ and there is no diffusion at all. Contrary, for the ratchet case,  $\alpha = 2$ and diffusion is ballistic indicating the wave-like phase motion and revealing an \emph{entirely new mechanism  responsible for anomalous diffusion}. This one should be clearly distinguished from the well known that appears in disordered systems \cite{sancho2004,khoury2011,sune2013,simon2014}. The latter form of the potential manifests also in the fact that there is a finite window of temperature for which the motion is super and subdiffusive. One can alter the regime of diffusion in the SQUID by proper adjustment of thermal noise intensity $Q$. In the high temperature limit  diffusion is normal regardless of the potential symmetry.  This observation agrees with our naive intuition since then the impact of both the conservative $U'(x)$ and the time-dependent $a\cos{(\omega t)}$ forces is negligible in comparison to thermal noise. In panel (b) the same quantity is presented  as a function of the external magnetic flux $\tilde{\Phi}_e$ for two selected temperatures $Q$. The unique feature is the ability to control the type of phase diffusion covering normal and anomalous regimes: from subdiffusion, to superdiffusion and finally the ballistic motion just by experimentally doable variation of $\tilde{\Phi}_e$. Moreover, in agreement with our previous statement we note that the phase diffusion regime is very sensitive to changes of this parameter and observe the rapid variability of $\alpha$ in the vicinity of $\tilde{\Phi}_e = 0$ and $\tilde{\Phi}_e = \pi/2$.
\subsection{Mechanism for transient anomalous diffusion}
\begin{figure}[t]
    \centering
    \includegraphics[width=1\linewidth]{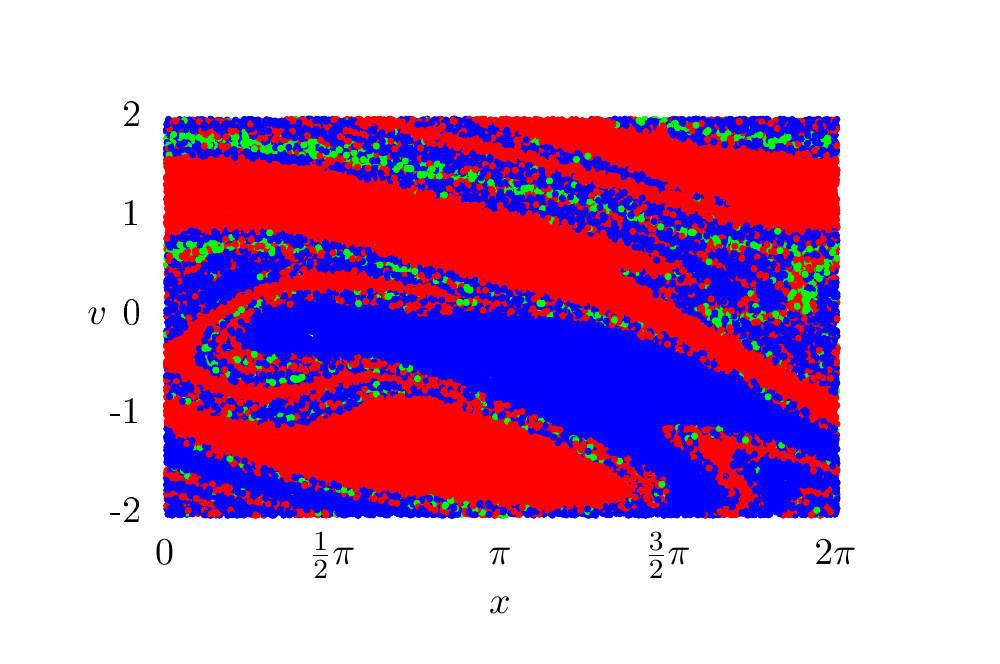}\\
    \includegraphics[width=0.9\linewidth]{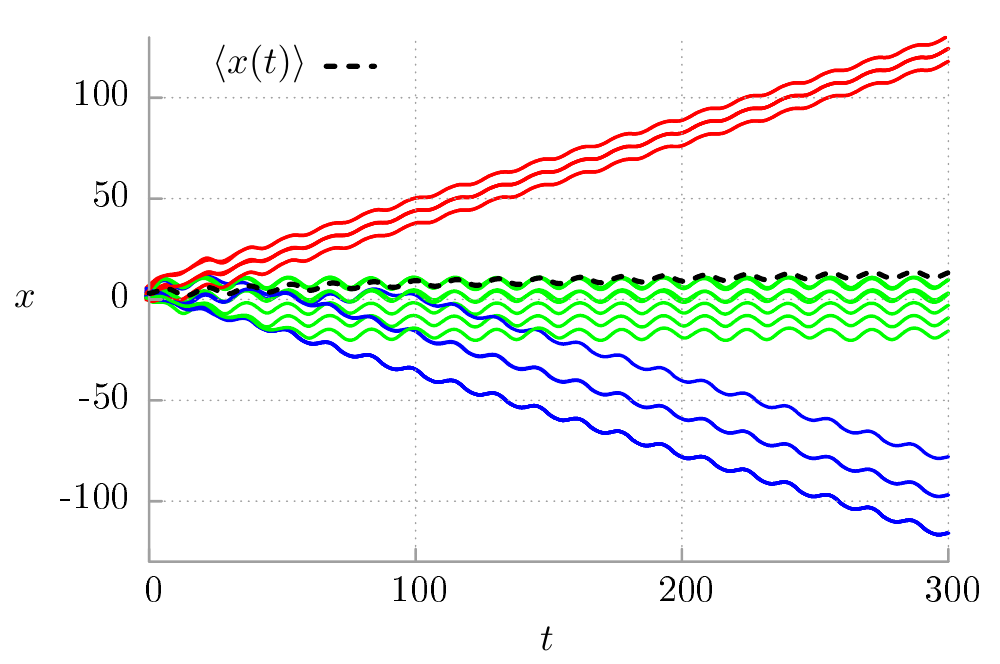}
    \caption{Basins of attraction for the asymptotic long time phase velocity $\langle v \rangle$ averaged over the period of the external ac driving $a\cos{\omega t}$ are presented in the upper panel. The bottom one depicts a number of sample realizations of the phase motion together with its ensemble averaged trajectory. Parameters are the same as in Fig. \ref{fig3} except the thermal noise intensity which is fixed to zero $Q=0$.}
    \label{fig6}
\end{figure}

To gain insight into the origin of transient anomalous diffusion, let us consider  the deterministic limit of vanishing thermal noise intensity $Q \to 0$  and  study the structure of basins of attraction for the asymptotic long time phase velocity $v = \dot{x}$ averaged over the period of the external ac driving, to be specific
\begin{equation}
 \langle v \rangle = \lim_{t \to \infty} \frac{\omega}{2\pi} \int_{t}^{t + 2\pi/\omega} ds\,\dot{x}(s). 
\end{equation}
The result is shown in the upper panel of Fig. \ref{fig6}. There exist only three attractors: the running state with either positive or negative phase velocity $\langle v \rangle = \pm 0.4$ (marked by red and blue color, respectively) or the locked state $\langle v \rangle = 0$ when the phase motion is limited to a finite number of potential wells (green color). The examples of corresponding trajectories are depicted in the bottom panel of the same figure together with the ensemble averaged temporal evolution of the phase across the device. This unexpected simplicity of attractors is crucial for the occurrence of the ballistic transport. In particular, qualitatively, due to the existence of two counter-propagating solutions with equal constant velocity $\langle v \rangle = 0.4$ the contribution of the average trajectory $\langle x(t) \rangle$ to the mean square displacement $\langle \Delta x^2(t) \rangle$ is vanishingly small in comparison to its second moment $\langle x^2(t) \rangle$, see the bottom panel of Fig. \ref{fig6}. As a consequence the mean square displacement is proportional to  time in the second power $\langle \Delta x^2(t) \rangle \sim t^2$. The application of thermal noise generally smooths out the complex structure of boundaries demarcating the coexisting attractors. In a qualitative picture, if temperature start to increase the phase is kicked out of its deterministic trajectory at random time and its mean value corresponds to the crossover time $\tau_1$. When temperature grows $\tau_1$ becomes to decrease what is exposed in Fig. \ref{fig3}.  Moreover, stochastic escape events among attractors give rise to other forms of the anomalous diffusion \cite{simon2014}.
\section{Temperature suppressed diffusion}
\begin{figure}[b]
    \centering
    \includegraphics[width=0.9\linewidth]{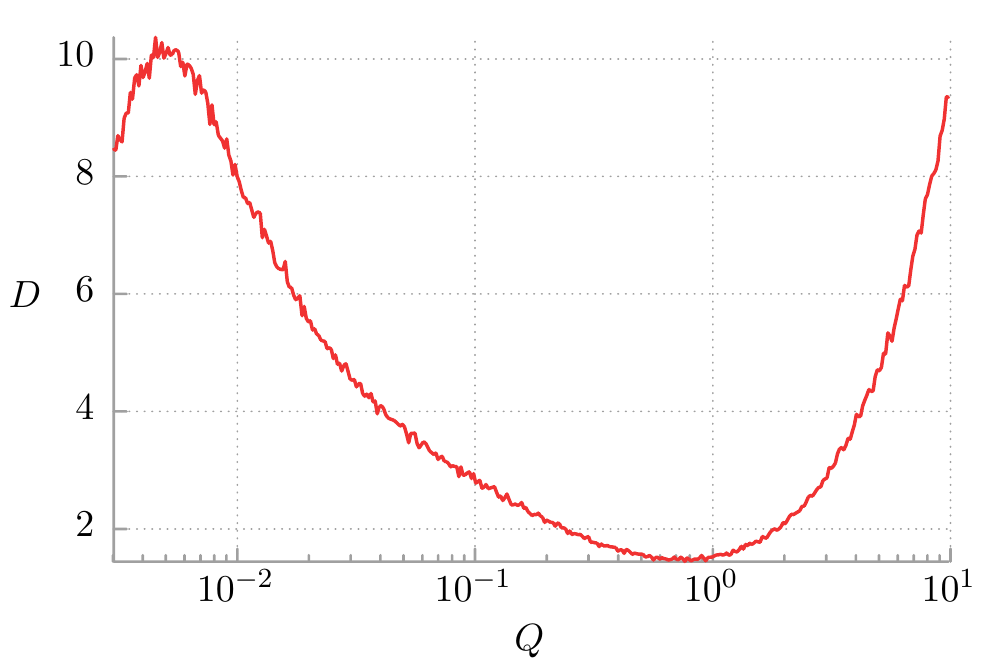}
    \caption{The diffusion coefficient $D$ versus the thermal noise intensity $Q$ in the high temperature limit. Other parameters are the same as in Fig. \ref{fig3}.}
    \label{fig7}
\end{figure}

Last but not least, let us analyze the diffusion coefficient $D$ in the normal diffusion regime. 
This scenario surely takes place in the limiting case of relatively high temperature $Q > 0.002$ since in such a case for times longer than $\approx 10^5$ the diffusion coefficient $D$ does not change with time (see Fig. \ref{fig3}b and Fig. \ref{fig5}a). Then it has a well established physical interpretation and can be conveniently computed by use of formula (\ref{eq6}). In Fig. \ref{fig7} we present its dependence on temperature, $Q \propto T$. The striking feature is its non-linear and non-monotonic behavior. For low temperature, $D$ initially increases as $Q$ grows, passes through its local maximum and next starts to decrease reaching its local minimum at some characteristic temperature $Q_c$. For temperatures higher that $Q_c$ the diffusion coefficient is a monotonically increasing function of $Q$. This \emph{temperature suppressed diffusion} phenomenon is in clear contrast with the Einstein relation $D \propto T$ as well as with the other known formulas as e.g for a Brownian particle moving in a periodic potentials \cite{lifson,festa} and in a tilted periodic potentials (under additional presence of an external constant force) \cite{reimann2001,lindner}.
\section{Summary} 
In this work we have investigated diffusion processes in the archetypal model of an inertial Brownian ratchet. As a particular realization we picked the asymmetric SQUID device driven by the time periodic current and pierced by the external constant magnetic flux. We have found selected parameter regimes for which the MSD of the Josephson phase evolves in three distinct stages: initially as superdiffusion, next as \mbox{subdiffusion} and finally as normal diffusion in the asymptotic long-time limit. We have shown that crossover times separating these three stages can be controlled by temperature and the external magnetic flux. The latter parameter is especially useful for this purpose as these times are particularly sensitive to its changes. Despite the fact that these abnormal processes have only transient nature they last many order longer than characteristic time scales of the system and thus they are comfortably detectable experimentally. Moreover, we have studied the origin of the discussed anomalous diffusion behavior and revealed the entirely new mechanism of its emergence which is based on breaking of reflection symmetry of the potential. This effect is particularly evident for low temperature regimes and should be clearly contrasted with the one that operates in disordered systems. In particular, in the deterministic limit of vanishing thermal noise the origin of the ballistic phase diffusion lies in the existence of two counter-propagating attractors. Finally, we have studied also the opposite limiting scenario of high thermal noise intensity and presented new manifest of the fascinating interplay between nonlinearity and thermal fluctuations, namely the phenomenon of noise suppressed diffusion.  

Our findings can be corroborated experimentally with a wealth of physical systems outlined in the introduction. One of the most promising setup for this purpose is the asymmetric SQUID device which has already been constructed
\cite{sterck2005,sterck2009}. It is due to its good experimental control as well as easy detection of the discussed diffusion behavior in the power spectrum of the voltage fluctuations. Such an experiment will probably not be as spectacular as those based on the single particle tracking technique \cite{burov2011,jeon2011,barkai2012}, however, probably it will be much easier to perform and therefore we think that our work may suggest a completely new \emph{testing ground} for investigating anomalies in diffusion phenomena.

\section*{Acknowledgement}
This work was supported by the MNiSW program ”Diamond Grant” (J. S.) and NCN grant DEC-2013/09/B/ST3/01659 (J. {\L}.)


\begin{thebibliography}{99}
	\bibitem{hanggi100years} P. H\"{a}nggi and F. Marchesoni, Chaos \textbf{15}, 026101 (2005)
	\bibitem{gammaitoni1998} L. Gammaitoni, P. H\"{a}nggi, P. Jung and F. Marchesoni, Rev. Mod. Phys. \textbf{70}, 223 (1998)
	\bibitem{hanggi2009} P. H\"{a}nggi and F. Marchesoni, Rev. Mod. Phys. \textbf{81}, 387 (2009)
	\bibitem{reimann2001} P. Reimann, C. Van den Broeck, H. Linke, P. H\"{a}nggi, J. M. Rubi and A. P\'erez-Madrid, Phys. Rev. Lett. \textbf{87}, 010602 (2001)
	\bibitem{spiechowicz2014pre} J. Spiechowicz, P. H\"{a}nggi and J. {\L}uczka, Phys. Rev. E \textbf{90}, 032104 (2014)
	\bibitem{metzler2000} R. Metzler and J. Klafter, Phys. Rep. \textbf{339}, 1 (2000)
	\bibitem{klafter2005} J. Klafter and I. M. Sokolov, Physics World \textbf{18}, 29 (2005)
	\bibitem{metzler2014} R. Metzler, J. H. Jeon, A. Cherstvy and E. Barkai, Phys. Chem. Chem. Phys. \textbf{16}, 24128 (2014)
	\bibitem{zaburdaev2015} V. Zaburdaev, S. Denisov, J. Klafter, arXiv:1410.5100
	\bibitem{sancho2004} J. M. Sancho, A. M. Lacasta, K. Lindenberg, I. M. Sokolov and A. H. Romero, Phys. Rev. Lett. \textbf{92}, 250601 (2004)
	\bibitem{khoury2011} M. Khoury, A. M. Lacasta, J. M. Sancho and K. Lindenberg, Phys. Rev. Lett. \textbf{106}, 090602 (2011)
	\bibitem{sune2013} M. Sune, J. M. Sancho and K. Lindenberg, Phys. Rev. E \textbf{88}, 062105 (2013)
	\bibitem{simon2014} M. S. Simon, J. M. Sancho and K. Lindenberg, Eur. Phys. J. B \textbf{87}, 201 (2014)
	\bibitem{bronstein2009} I. Bronstein, Y. Israel, E. Kepten, S. Mal, Y. Shav-Tai, E. Barkai and Y. Garini, Phys. Rev. Lett. \textbf{103}, 018102 (2009)
	\bibitem{hanes2012theo} R. D. L. Hanes and S. U. Egelhaaf, J. Phys.: Condens. Matter \textbf{24}, 464116 (2012)
	\bibitem{hanes2012exp} R. D. L. Hanes, C. Dalle-Ferrier, M. Schmiedeberg, M. C. Jenkins and S. U. Egelhaaf, Soft Matter \textbf{8}, 2714 (2012)
	\bibitem{hanes2013} R. D. L. Hanes, M. Schmiedeberg and S. U. Egelhaaf, Phys. Rev. E \textbf{88}, 062133 (2013)
	\bibitem{machura2007} {\L.} Machura, M. Kostur, P. Talkner, J. {\L}uczka and P. H\"{a}nggi, Phys. Rev. Lett. \textbf{98}, 40601 (2007)
	\bibitem{speer2007} D. Speer, R. Eichhorn and P. Reimann, Europhys. Lett. \textbf{79}, 10005 (2007)
	\bibitem{nagel2008} J. Nagel, D. Speer, T. Gaber, A. Sterck, R. Eichhorn, P. Reimann, K. Ilin, M. Siegel, D. Koelle, and R. Kleiner, Phys. Rev. Lett. \textbf{100}, 217001 (2008)
	\bibitem{kostur2008} M. Kostur, {\L}. Machura, P. Talkner, P. H\"{a}nggi and J. {\L}uczka, Phys. Rev. B, \textbf{77}, 104509 (2008)
	\bibitem{karnik2007} R. Karnik, C. Duan, K. Castelino, H. Dalguji and A. Majumdar, Nano Lett. \textbf{7}, 547 (2007)
	\bibitem{gommers2005} R. Gommers, S. Bergamini and F. Renzoni, Phys. Rev. Lett. \textbf{95}, 073003 (2005)
	\bibitem{arzola2011} A. Arzola, K. Volke-Sep\'{u}lveda and J. L. Mateos, Phys. Rev. Lett. \textbf{106}, 168104 (2011)
	\bibitem{denisov2014} S. Denisov, S. Flach and P. H\"{a}nggi, Phys. Rep. \textbf{538}, 77 (2014)
	\bibitem{lee1999} C. S. Lee, B. Jank\'{o}, I. Der\'{e}nyi and A. L. Barab\'{a}si, Nature \textbf{400}, 337 (1999)
	\bibitem{villegas2003} J. E. Villegas, S. Savel'ev, F. Nori, E. M. Gonzalez, J. V. Anguita, R. Arcia and J. L. Vincent, Science \textbf{302}, 1188 (2003)
	\bibitem{ustinov2004} A. V. Ustinov, C. Coqul, A. Kemp, Y. Zolotaryuk and M. Salerno, Phys. Rev. Lett. \textbf{93}, 087001 (2004)
	\bibitem{beck2005} M. Beck, E. Goldobin, M. Neuhaus, M. Siegel, R. Kleiner and D. Koelle, Phys. Rev. Lett. \textbf{95}, 090603 (2005)
	\bibitem{knufinke2012} M. Knufinke, K. Ilin, M. Siegel, D. Koelle, R. Kleiner and E. Goldobin, Phys. Rev. E \textbf{85}, 011122 (2012)
	\bibitem{weiss2000} S. Weiss, D. Koelle, J. M\"{u}ller, R. Gross and K. Barthel, Europhys. Lett. \textbf{51}, 499 (2000)
	\bibitem{sterck2005} A. Sterck, R. Kleiner and D. Koelle, Phys. Rev. Lett. \textbf{95}, 177006 (2005)
	\bibitem{sterck2009} A. Sterck, D. Koelle and R. Kleiner, Phys. Rev. Lett. \textbf{103}, 047001 (2009)
	\bibitem{zapata1996} I. Zapata, R. Bartussek, F. Sols and P. H\"{a}nggi, Phys. Rev. Lett. \textbf{77}, 2292 (1996)
	\bibitem{spiechowicz2014} J. Spiechowicz, P. H\"{a}nggi and J. {\L}uczka, Phys. Rev. B \textbf{90}, 054520 (2014)
	\bibitem{renz12} A. Wickenbrock, P. C. Holz,  N. A. Abdul Wahab,  P. Phoonthong,  D. Cubero,  and F. Renzoni,  Phys. Rev. Lett. \textbf{108}, 020603 (2012). 
	\bibitem{renz13} E. Lutz and F. Renzoni, Nat. Phys. \textbf{9}, 615 (2013). 
	\bibitem{atbook} R. Klages, G. Radons, and I. M. Sokolov, \textit{Anomalous transport: Foundations and applications} (Wiley-VCH, Weinheim, 2008)
	\bibitem{geisel} T. Geisel, J. Nierwetberg and A. Zacherl, Phys. Rev. Lett. \textbf{54}, 616 (1985)
	\bibitem{ishizaki2003} R. Ishizaki, S. Kuroki, H. Tominaga, N. Mori and H. Mori, Prog. Theor. Phys. \textbf{109}, 169 (2003)
	\bibitem{jung1993} P. Jung, Phys. Rep. \textbf{234}, 175 (1993). 
	\bibitem{jung1996} P. Jung, J. G. Kissner and P. H\"{a}nggi, Phys. Rev. Lett. \textbf{76}, 3436 (1996)
	\bibitem{mateos2000} J. L. Mateos, Phys. Rev. Lett. \textbf{84}, 258 (2000)
	\bibitem{spiechowiczcpc} J. Spiechowicz, M. Kostur and {\L}. Machura, Comp. Phys. Commun. \textbf{191}, 140 (2015) 
	\bibitem{NJP15} J. Spiechowicz and J. {\L}uczka, New. J. Phys. \textbf{17}, 023054 (2015).
	\bibitem{lifson} S. Lifson and J. L. Jackson, J. Chem. Phys. 36, 2410 (1962).
	\bibitem{festa} R. Festa and E. Galleani d'Agliano, Physica A 90, 229 (1978).
	\bibitem{lindner} B. Lindner, M. Kostur, and L. Schimansky-Geier, Fluct. Noise Lett. 1, R25 (2001).
	\bibitem{burov2011} S. Burov, J. H. Jeon, R. Metzler, E. Barkai, Phys. Chem. Chem. Phys. \textbf{13}, 1800 (2011)
	\bibitem{jeon2011} J. H. Jeon, V. Tejedor, S. Burov, E. Barkai, Ch. Selhuber-Unkel, K. Berg-Sorensen, L. Oddershede and R. Metzler, Phys. Rev. Lett. \textbf{106}, 048103 (2011)
	\bibitem{barkai2012} E. Barkai, Y. Garini, R. Metzler, Physics Today \textbf{65}, 29 (2012)
\end{thebibliography}
\end{document}


\title{Addendum to "Diffusion anomalies in ac-driven Brownian ratchets"}
\author{Jakub Spiechowicz}
\affiliation{Institute of Physics, University of Silesia, 40-007 Katowice, Poland}
\affiliation{Silesian Center for Education and Interdisciplinary Research, University of Silesia, 41-500 Chorz{\'o}w, Poland}
\author{Jerzy {\L}uczka}
\affiliation{Institute of Physics, University of Silesia, 40-007 Katowice, Poland}
\affiliation{Silesian Center for Education and Interdisciplinary Research, University of Silesia, 41-500 Chorz{\'o}w, Poland}
\email{jerzy.luczka@us.edu.pl}

%
\begin{abstract}
The origin of diffusion anomalies discussed in our  previous work [Phys. Rev. E \textbf{91}, 062104 (2015)] was related to the \emph{ratchet} form of the periodic structure thus revealing a new mechanism of emergence of anomalous diffusion. In this addendum we clarify the role of breaking of the potential reflection symmetry in the observed phenomenon.
\end{abstract}
\pacs{
05.10.Gg, 
05.40.-a, 
05.40.Ca, 
05.60.-k, 
85.25.Cp, 
85.25.Dq 
}
\maketitle
\begin{figure}[t]
	\centering
	\includegraphics[width=1\linewidth]{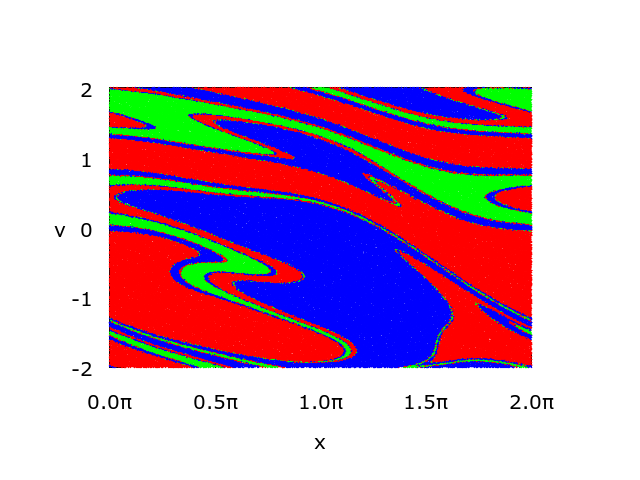}
	\caption{Basins of attraction for the asymptotic long-time phase velocity $\langle v \rangle$ averaged over the temporal period $2\pi/\omega$ of the external driving $a\cos{(\omega t)}$: $\langle v \rangle = 0.4$ (red), $\langle v \rangle = -0.4$ (blue), $\langle v \rangle \approx 0 $ (green). Parameters are the same as in Fig. 3 of Ref. [1] except now the potential $U(x)$ is \emph{symmetric} with $\tilde{\Phi}_e = \pi$.}
	\label{fig1}
\end{figure}
\begin{figure}[b]
	\centering
	\includegraphics[width=1\linewidth]{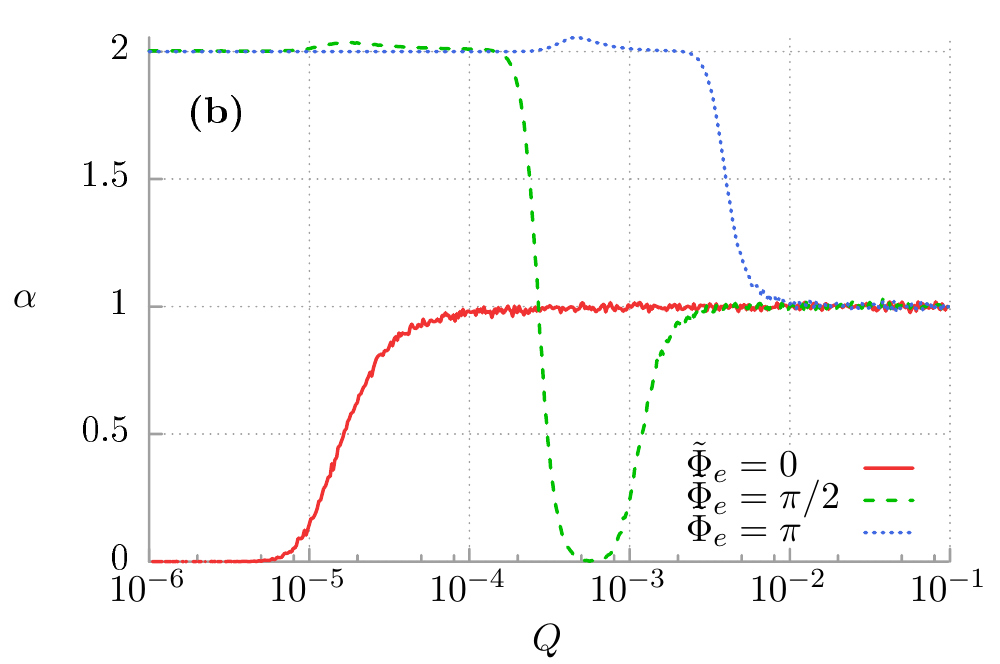}
	\caption{The power exponent $\alpha$ versus temperature $Q \propto T$ for selected values of the external magnetic flux $\tilde{\Phi}_e$. $\tilde{\Phi}_e = 0$ and $\tilde{\Phi}_e = \pi$ corresponds to the symmetric potential $U(x)$ whereas for $\tilde{\Phi}_e = \pi/2$ it is asymmetric. Other parameters are the same as in Fig. 3 of Ref. \cite{spiechowicz2015pre}.}
	\label{fig2}
\end{figure}

In Sec. III.C of Ref. \cite{spiechowicz2015pre}, we comment on the power exponent $\alpha$ characterizing a type of diffusion. 
In Fig. 5 (a) there,  the emergence of superdiffusion (ballistic diffusion)  with $\alpha=2$  is illustrated in the low temperature regime (small values of the thermal noise intensity $Q$) for the ratchet potential $U(x)$ with $\tilde{\Phi}_e = \pi/2$. This behaviour was  associated with the existence of two counter-propagating attractors in the corresponding deterministic system (see Fig. 6 in [1]) and next contrasted with the case when the potential is reflection symmetric for   $\tilde{\Phi}_e = 0$.  
We wrote: 'When potential is reflection
symmetric, then $\alpha \approx 0$  and there is no diffusion at all'. It is 
accurate  when $\tilde{\Phi}_e = 0$. However, the potential is symmetric also for $\tilde{\Phi}_e = \pi$ and in this case $\alpha =2$ what can be directly inferred from Fig. 5 (b) in \cite{spiechowicz2015pre}. 
 Now, we \emph{clarify} that the special structure of attractors for the ratchet with $\tilde{\Phi}_e = \pi/2$ can also be observed for the \emph{symmetric} potential with $\tilde{\Phi}_e = \pi$ (see Fig. 1) leading, in agreement with the mechanism outlined in \mbox{Ref. \cite{spiechowicz2015pre}}, to the ballistic diffusion with $\alpha =2$.   
 However, the detected \emph{sequence} of superdiffusion - subdiffusion - normal diffusion is observed in this regime \emph{only} when reflection symmetry of the potential $U(x)$ is broken, i.e. when it has a ratchet form. We support this statement by presenting  the explicit dependence of the power exponent $\alpha$ on thermal noise intensity $Q \propto T$ for selected values of the external magnetic flux $\tilde{\Phi}_e = 0, \pi/2$ and $\pi$,  see Fig. \ref{fig2}. 

\begin{figure}[t]
	\centering
	\includegraphics[width=1\linewidth]{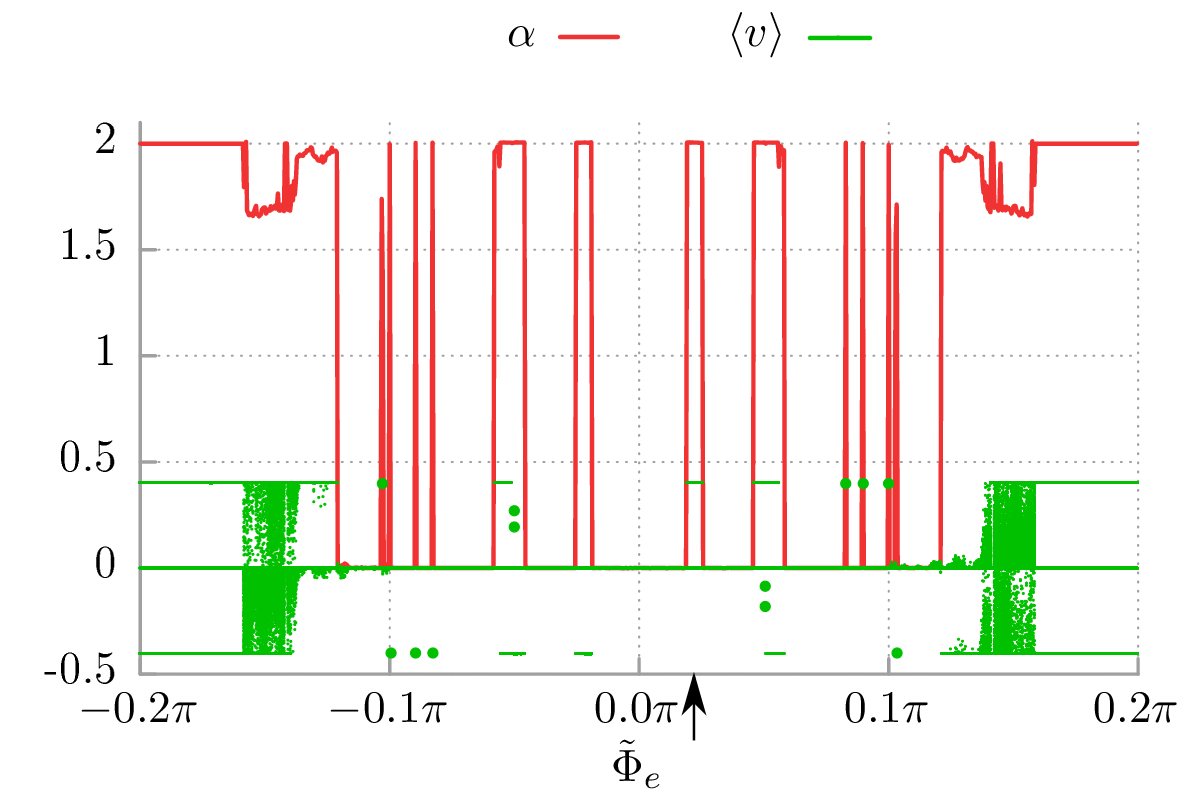}
	\caption{The power exponent $\alpha$ and the asymptotic long time velocity $\langle v \rangle$ averaged over the temporal period $2\pi/\omega$ of the external driving  $a\cos{(\omega t)}$ depicted versus the external magnetic flux $\tilde{\Phi}_e$ in the deterministic limit $Q = 0$. Further increase of \mbox{$|\tilde{\Phi}_e| > 0.2\pi$} does not introduce any significant changes in neither the power exponent $\alpha$ nor the asymptotic velocity $\langle v \rangle$. The arrow indicates $\tilde{\Phi}_e \approx 0.022\pi$. Other parameters are the same as in Fig. 3 of Ref. \cite{spiechowicz2015pre}.}
	\label{fig3}
\end{figure}
To establish connection between the deterministic structure of attractors and the discussed diffusion anomalies we now study the corresponding bifurcation diagram illustrating all stable orbits for the asymptotic long time velocity $\langle v \rangle$ averaged over the temporal period $2\pi/\omega$ of the external driving $a\cos{(\omega t)}$. The result is shown in Fig. \ref{fig3}. Indeed, we observe that in each case when in the deterministic limit there are two regular attractors transporting in the opposite direction with the same magnitude,  the power exponent $\alpha = 2$. We note that diffusion is ballistic also when there is only one locked attractor $\langle v \rangle \approx 0$ accompanied by the running solution $|\langle v \rangle| \approx 0.4$ with a much smaller population of the initial conditions (e.g. for $\tilde{\Phi}_e \approx 0.022\pi$ depicted  by the arrow in Fig. 3). The latter fact can be directly seen from the corresponding basins of attraction which is not depicted here. On the contrary, there is no diffusion at all when in the long time there exists only one stable orbit since then eventually all trajectories will follow it. Overall, when in the deterministic limit there are several running solutions associated with different velocities then it is difficult to define a type of diffusion \emph{a priori}. The relative size of the corresponding basins of attraction can be helpful in this task. Then the diffusive behaviour may be either normal or anomalous \cite{spiechowicz2015chaos}.

In summary, we clarified the role of breaking of the potential reflection symmetry in the observed diffusion anomalies. The \emph{sequence} of superdiffusion-subdiffusion-normal diffusion is detected in the discussed regime \emph{only} when the potential has a \emph{ratchet} form. However, the structure of counter-propagating attractors may emerge also for \emph{symmetric} periodic systems immediately implying superdiffusion in low temperature regimes. 

\section*{Acknowledgement}
This work was supported by the MNiSW program ”Diamond Grant” (J. S.).